# Liver Pathology Simulation: Algorithm for Haptic Rendering and Force Maps for Palpation Assessment

Felix G. HAMZA-LUP[a], Adrian SEITAN[b], Dorin M. POPOVICI[b], and Crenguta M. BOGDAN[b]

[a] *Computer Science and Information Technology, Armstrong Atlantic State University, Savannah, Georgia, US*
[b] *Mathematics and Informatics Faculty, Ovidius University, Constanta*

**Abstract.** Pre-operatory gestures include tactile sampling of the mechanical properties of biological tissue for both histological and pathological considerations. Tactile properties used in conjunction with visual cues can provide useful feedback to the surgeon. Development of novel cost effective haptic-based simulators and their introduction in the minimally invasive surgery learning cycle can absorb the learning curve for your residents. Receiving pre-training in a core set of surgical skills can reduce skill acquisition time and risks. We present the integration of a real-time surface stiffness adjustment algorithm and a novel paradigm - *force maps* - in a visuo-haptic simulator module designed to train internal organs disease diagnostics through palpation.

**Keywords.** haptic rendering, simulation, laparoscopy, liver pathology.

## Introduction

Surgery has always been a tactile intensive activity and the tactile perception plays an important role in surgery. The surgeon must feel organic tissue hardness, evaluate anatomical structures, measure tissue properties, and apply appropriate force control actions for safe tissue manipulation.

Unfortunately most East-European countries use obsolete and sometimes financially inefficient training systems. Procedures are practiced on live animals [1] and there is only one center with computer based simulators [2] in the entire region. In an effort to improve laparoscopic training we developed a simple yet cost-effective environment for surgical training using of-the-shelf hardware components. We focused on simulating liver related pathologies since the liver is one of the most affected organs. Hepatitis C virus infection is a growing public health concern. Globally an estimated 180 million people, or roughly 3% of the world's population, are currently infected [3].

We present a liver palpation force assessment method using a novel concept: dynamic force maps and an improvement on the haptic rendering algorithm available under the H3D platform [4] for heterogeneous stiffness computation.

### 1. Liver Tissue – Heterogeneous Stiffness Simulation

Tissue modeling holds a very important place in medical simulation as the realism of the simulation relies on the soft tissue response during the simulated intervention. Elastic and plastic deformations of 3D models have been extensively researched in the past decade. Among the most notable efforts we mention the General Physical Simulation Interface (GiPSi) [5], an open-source framework that presents a flexible architecture, developed to simulate surgical procedures at organ level. Simulation Open Framework Architecture (SOFA) [6] an international, multi-institution, collaborative initiative, aimed at developing a flexible and open source framework for interactive simulations. Computer Haptics and Active Interfaces - CHAI3D [7] an open-source designed to facilitate the development of 3D modeling applications augmented with haptic rendering. Other attempts at designing software toolkits for medical simulation are SPORE [8], SSTML [9] and Spring [10]. A review of the APIs, frameworks and toolkits for haptic-based simulation of soft tissue deformation can be found in [11].

Numerical methods for approximate solutions to partial differential equations like the Finite Element Method (FEM) have gained some ground in interactive simulations. FEM is computationally intensive however recently interactive speeds were obtained as the computation is parallelized on the GPU [12].

A more generic API that offers soft tissue modeling algorithms is Haptics3D (H3D) [4]. The main advantages of H3D are the rapid prototyping capability and the compatibility with eXtended 3D (X3D), making it easy for the developer to manage both the 3D graphics and the haptic rendering. H3D API uses the X3D and OpenGL standards and builds on haptic technology from SensAble's OpenHaptics™ toolkit [13].

Two important factors when implementing deformable 3D tissue are the tissue's surface properties and its stiffness. The first one defines the visual and haptic properties at touch while the second one is part of the deformation algorithm that provides visual as well as force feedback during tissue interaction (e.g., palpation, cutting).

### 2. RGB Color Maps

We propose a simple yet effective method to simulate heterogeneous stiffness properties along the liver soft tissue: RGB color maps. Our algorithm improves upon the H3D haptic rendering algorithms. We use two geometries, one for the visual rendering and one for the haptic rendering.

The visual deformation of the tissue depends on the location and the force applied on the surface. The visual deformation algorithm under the H3D platform uses a Gauss function of form Eqn (2), where *a* represents the amplitude of the Gauss function, *w* controls the width of the kernel, *diff* represents the penetration distance inside the soft tissue and *e* is the Euler's number.

$$f(x) = a * e^{\frac{-(diff * diff)}{w * w}} \tag{1}$$

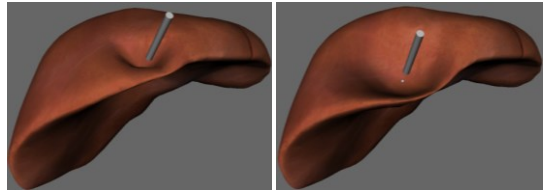

**Figure 1.** Left: Stiffness=0.2, width=0.02, amplitude=1. Right: Stiffness=0.2, width=0.05, amplitude=1.2

The force feedback is computed from the haptic model considering the penetration distance $\Delta x$, an elasticity coefficient $k$ as well as two parameters from the deformation node: stiffness – representing the stiffness value of the model in that vertex and damping – representing the spring damping factor associated with the vertex.

$$F = \Delta x * k \tag{2}$$

A higher damping factor means that the elasticity of the tissue in that point is reduced. While the impact of the damping factor cannot be seen in a static image, Figure 1 illustrates the impact of the kernel width $w$ and the amplitude $a$ on the localized visual deformation of the 3D model.

Building on the above haptic rendering model we use the RGB values associated with each vertex of the polygonal model to store values that will be later haptically interpreted as tissue stiffness properties based on the pathology (e.g., the location, size and stiffness of cystic tissue). Each vertex RGB values are normalized on [0,1] interval and instead of being interpreted as colors, they will be interpreted as stiffness values. The stiffness values are generated during the 3D modeling phase according to the CT scans associated with a specific patient. Since we are actually generating 3D models for a simulator, we generated multiple liver models with various disease conditions. These models are used in the simulator to train students to diagnose liver pathologies through palpation.

The set of stiffness values can be seen as a stiffness map (Figure 2) that will be interpreted during the haptic rendering process, allowing heterogeneous stiffness properties simulation along the liver surface.

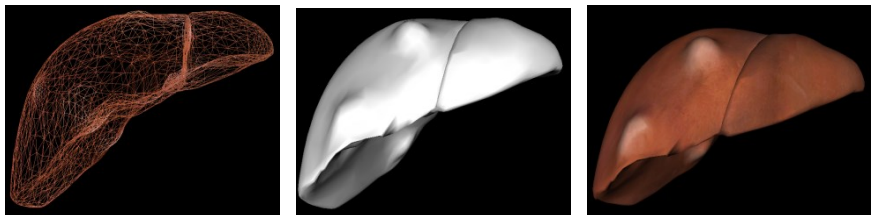

**Figure 2.** The wire-frame (left), solid (center) and textured (right) models.

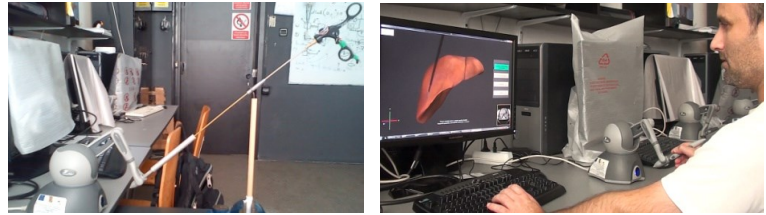

**Figure 3.** Simulator prototype with trocar insertion (left), direct manipulation (right).

## 3. Liver Pathology Simulation

Liver palpation can reveal multiple issues: presence of emphysema with an associated depressed diaphragm, fatty infiltration (enlarged with rounded edge), active hepatitis (enlarged and tender), cirrhosis (enlarged with nodular irregularity), and hepatic neoplasm (enlarged with rock-hard or nodular consistency).

The liver pathology simulator (Figure 3) is a cost-effective assessment tool that represents different liver pathologies combined with sets of questionnaires to verify the student theoretical and practical knowledge accumulated in the field so far.

### 3.1. Healthy and Hepatic Liver

The healthy liver model started from a large 3D polygonal model originated from an anonymized CT scan. The polygonal model has been decimated up to a certain threshold (3K polygons) to improve the rendering speed while maintaining visual quality (Figure 4). The decimation threshold was determined empirically testing increasingly lower resolutions while collecting feedback from our medical team.

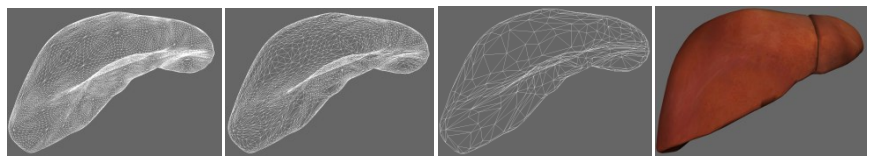

**Figure 4.** Model, level of detail: 13K , 6K, 3K, 3K+texture

The Hepatic liver case uses a similar geometry but a different texture as well as slightly different stiffness properties. Students must be capable of recognizing a hepatic liver starting from its visual clues, texture and tissue coloration, and then using the haptic device.

### 3.2. Cirrhosis

Cirrhosis is a consequence of chronic liver disease characterized by replacement of liver tissue by fibrosis, scar tissue and regenerative nodules (lumps that occur as a result of a process in which damaged tissue is regenerated) [14] leading to loss of liver function. Cirrhosis is most commonly caused by alcoholism, hepatitis B and C, and fatty liver disease, but has many other possible causes. To simulate the cirrhotic liver tissue we used 3D sculpting modeling methods and the 3DMax platform [15] to alter the polygonal mesh. The haptic rendering was adjusted mainly in the stiffness and damping parameters based on feedback from the surgeons collaborators.  Figure 5 illustrates interaction of the Babcock pense with the cirrhotic liver model. Visual rendering is executed at a minimum of 59 frames per second (FPS) while the haptic rendering cycle is maintained close to 1 KHz at 997 Hz.

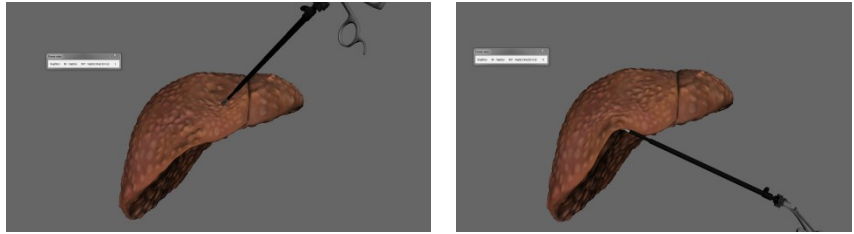

**Figure 5.** Cirrhotic liver simulation.

*3.3. Liver Cysts and Cystic Tumors*

Liver cysts occur in approximately 5% of the population. However, only about 5% of these patients ever develop symptoms. In general, cysts are thin-walled structures that contain fluid. Most cysts are single, although some patients may have several. Unlike simple liver cysts, cystic tumors are actually growths that may become malignant over the course of many years. The benign cystic tumor seen most frequently is called a cystadenoma; its malignant counterpart is a cystadenocarcinoma. CT scans are the best imaging studies to show the cystic tumors, which contain both liquid and solid areas [16].

To haptically represent the cysts we modified the stiffness and damping factors in the cyst region using the RGB color maps algorithm. The user receives additional haptic feedback, based on stiffness variation in the regions where deep cysts exist but are not visible at the surface.  Figure 6 illustrates this idea highlighting with a dark spot the cystic region that will "feel" differently during haptic interaction but look visually the same as other parts of the liver tissue.

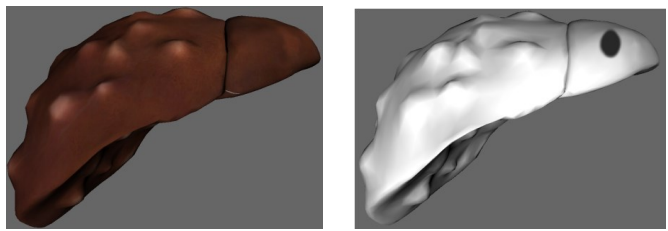

**Figure 6.** Left: 3D model of the cystic liver. Right: same model with RGB codes, the dark spot will be rendered only haptically (prototype low resolution models)

**4. Palpation Force Assessment**

The force applied during palpation must be maintained in a certain range. Palpation with small forces may not reveal correctly mechanical properties of the biological tissue, while forces exceeding a certain threshold can damage healthy tissue.

*4.1. Interactive Palpation Force Measurement*

We implemented a dynamic force measurement approach and a visualization module to find the appropriate range of forces during the liver palpation procedure collecting force data directly from the experienced surgeons. The module draws a force measurement indicator range on the left side of the screen (Figure 7). The range empirically agreed upon is in the interval 2.1 to 2.5 Newtons. A standard Babcock pense was connected to the Phantom Omni device [13] to practice palpation.

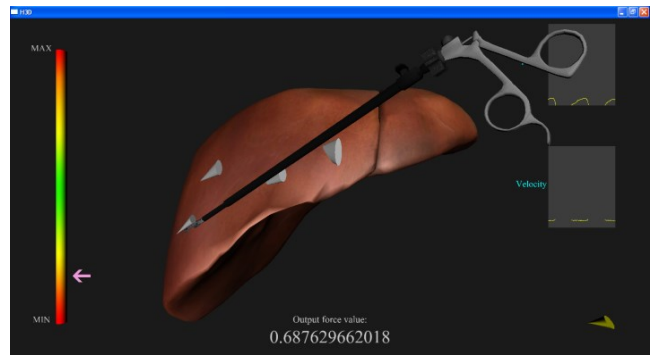

**Figure 7.** Force measurement and range estimation

*4.2. Pense Force Dynamics and Surface Velocity*

Other interaction attributes that we consider are the palpation force dynamics and the pense surface velocity. During the actual simulation, the prototype records several simulation parameters (position, force, velocity etc.) at a frequency of 100Hz.

*4.3. Force Map Visualization*

The prototype represents the palpation force, position and orientation thought cones directly on the liver's surface. The cone's height and bottom radius are proportional with the magnitude of the force applied on the tissue's surface. Moreover the position and orientation of the pense is represented by the cone's height direction. So the evaluator can see not only the force applied but also the location and the direction of the pense relative to the liver surface. The assessment method takes into consideration the palpation gesture according to the type of liver.

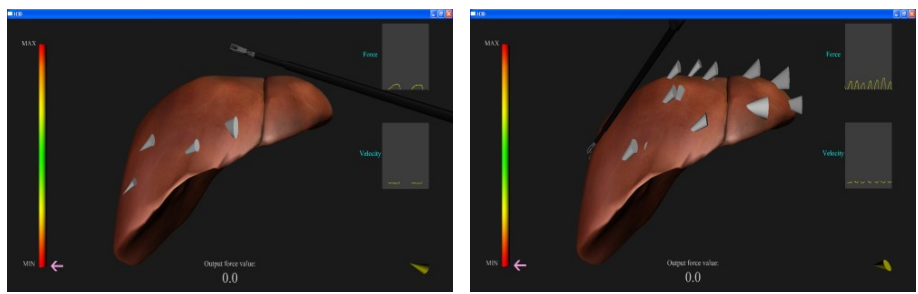

**Figure 8.** Experienced surgeon (left). Novice student (right)

In Figure 8 (left) the user is an experienced surgeon: the palpation force used on each "tap" on the liver's surface is constant. We observe that the velocity of the Babcock pense on the liver surface is constant too. Figure 8 (right) the user is a novice: the palpation force and the haptic device's velocity varies abruptly when it should remain at a relative constant value to avoid tissue damage.

**Conclusions**

We have presented a simple yet efficient algorithm improvement for heterogeneous stiffness simulation combined with a palpation force assessment module. While large scale assessment is in progress, we have already observed patterns for the Babcock pense force dynamics and surface velocity that can differentiate a novice from an expert surgeon on several levels.

**Acknowledgments**
This study was supported under the ANCS grant "HapticMed – Using haptic interfaces in medical applications", no. 128/02.06.2010, ID/SMIS 567/12271.